\documentclass[conference]{IEEEtran}

\usepackage{cite}
\usepackage{amsmath,amssymb,amsfonts}
\usepackage{graphicx}
\usepackage{textcomp}
\usepackage{xcolor}
\usepackage{soul}
\usepackage{tabularx}
\usepackage{multirow}
\usepackage{cite}
\usepackage{hyperref}
\usepackage{comment}

\usepackage{algpseudocode}
\usepackage{algorithm}
\usepackage{xspace}
\usepackage{amsmath}
\usepackage{upgreek}
\newtheorem{definition}{Definition}

\usepackage{threeparttable}  
\usepackage{booktabs}
\usepackage{multirow}
\usepackage{bm}

\algdef{SE}[DOWHILE]{Do}{doWhile}{\algorithmicdo}[1]{\algorithmicwhile\ #1}%

\usepackage{tikz}
\usetikzlibrary{arrows.meta,positioning}

 \newcommand{\Aa}{\ensuremath{{\mathcal{A}_{I}}}}
 
 \newcommand{\Ab}{\ensuremath{{\mathcal{A}_{II}}}}
 
 \newcommand{\Ac}{\ensuremath{{\mathcal{A}_{III}}}}
 
 \newcommand{\Ad}{\ensuremath{{\mathcal{A}_{IV}}}}

\usepackage{etoolbox}
\newcommand{\circled}[2][]{%
	\tikz[baseline=(char.base)]{%
		\node[shape = circle, draw, fill=red, color=red, inner sep = .2pt]
		(char) {\phantom{\ifblank{#1}{#2}{#1}}};%
		\node at (char.center) {\makebox[0pt][c]{\color{white}{#2}}};}}
\robustify{\circled}

\def\BibTeX{{\rm B\kern-.05em{\sc i\kern-.025em b}\kern-.08em
    T\kern-.1667em\lower.7ex\hbox{E}\kern-.125emX}}

\title{RAG Security and Privacy: Formalizing the Threat Model and Attack Surface}

 \makeatletter
\newcommand{\linebreakand}{%
  \end{@IEEEauthorhalign}
  \hfill\mbox{}\par
  \mbox{}\hfill\begin{@IEEEauthorhalign}
}
\makeatother
 \newcommand{\linebreakanda}{%
   \end{@IEEEauthorhalign}
   \hfill\mbox{}\par
   \mbox{}\hfill\begin{@IEEEauthorhalign}
 }
 \author{\IEEEauthorblockN{Atousa Arzanipour}
 \IEEEauthorblockA{
 University of South Florida\\
      Tampa, USA \\
     arzanipour@usf.edu}

 \and
 \IEEEauthorblockN{Rouzbeh Behnia}
 \IEEEauthorblockA{
 University of South Florida\\
    Tampa, USA \\
  behnia@usf.edu}
 \and
 \IEEEauthorblockN{Reza Ebrahimi}
 \IEEEauthorblockA{
 University of South Florida\\
    Tampa, USA \\
  ebrahimim@usf.edu}
 \and
 \IEEEauthorblockN{Kaushik Dutta}
 \IEEEauthorblockA{
 University of South Florida\\
    Tampa, USA \\
  duttak@usf.edu}

 }
    
\IEEEoverridecommandlockouts

\begin{document}

\maketitle
\thispagestyle{plain}
\pagestyle{plain}

\begin{abstract}
Retrieval-Augmented Generation (RAG) is an emerging approach in natural language processing that combines large language models (LLMs) with external document retrieval to produce more accurate and grounded responses. While RAG has shown strong potential in reducing hallucinations and improving factual consistency, it also introduces new privacy and security challenges that differ from those faced by traditional LLMs. Existing research has demonstrated that LLMs can leak sensitive information through training data memorization or adversarial prompts, and RAG systems inherit many of these vulnerabilities. At the same time, RAG’s reliance on an external knowledge base opens new attack surfaces, including the potential for leaking information about the presence or content of retrieved documents, or for injecting malicious content to manipulate model behavior. Despite these risks, there is currently no formal framework that defines the threat landscape for RAG systems. In this paper, we address a critical gap in the literature by proposing, to the best of our knowledge, the first formal threat model for retrieval-RAG systems. We introduce a structured taxonomy of adversary types based on their access to model components and data, and we formally define key threat vectors such as document-level membership inference and data poisoning, which pose serious privacy and integrity risks in real-world deployments. By establishing formal definitions and attack models, our work lays the foundation for a more rigorous and principled understanding of privacy and security in RAG systems.  
\end{abstract}


\section{Introduction}

Large Language Models (LLMs) are trained on a vast amount of data with the goal of understanding and generating human language \cite{devlin2018bert}. They are widely applied across different domains, such as chatbots. However, the issue with LLMs is that their performance is constrained by the quality and scope of their training data. For instance, if a chatbot is asked a real-time or domain-specific question, its supporting LLM may generate an answer that seems logical but is actually incorrect, a phenomenon known as hallucination.

Retrieval-Augmented Generation (RAG) is an emerging paradigm in natural language processing and generative AI that combines the generative capabilities of LLMs with dynamic access to external knowledge repositories \cite{lewis2020retrieval}. In a typical RAG architecture, a retriever first identifies relevant documents or passages from a knowledge base or online sources based on an input query, and a generator then conditions on both the query and the retrieved content to produce a response \cite{izacard2020leveraging}. By integrating retrieval into the generation process, RAG systems enhance the factual accuracy \cite{joren2025sufficient}, coherence \cite{yi2025score}, and contextual grounding of generated outputs across a range of applications. This integration has proven particularly beneficial in complex tasks such as fact-checking \cite{khaliq2024ragar} and information retrieval \cite{izacard2020leveraging}, where accurate and timely access to external information is critical.

The practical impact of RAG is increasingly evident in industrial settings: major search engines such as Google Search and Microsoft Bing are exploring the incorporation of RAG-based systems to improve their response quality by leveraging both curated knowledge bases and real-time web content \cite{sato2024google,steen2024bing}. By mitigating issues like hallucination \cite{ji2023survey}, factual inconsistency, and limited generalization \cite{misrahi2025adapting} that often affect standalone LLMs, RAG architectures offer a promising foundation for building more reliable and knowledge-aware AI systems.

While RAG systems offer improvements in factual accuracy and contextual grounding, their foundation, LLM, remains susceptible to a range of privacy and security threats that arise during both training and inference. During training, LLMs can memorize and inadvertently expose sensitive data from the training corpus \cite{carlini2023extracting}. At inference time, adversaries may exploit vulnerabilities such as prompt injection \cite{Greshake2023}, model misalignment \cite{pandey2025accidental}, or gradient inversion \cite{zhang2022survey}, leading to biased, harmful, or unintended outputs. These threats undermine the reliability, fairness, and safety of LLM-based systems, particularly in sensitive or regulated domains (e.g., healthcare).

RAG systems are built on top of LLMs and therefore inherit these vulnerabilities. In addition, they introduce a new class of privacy and security risks due to their architectural design. Unlike conventional LLMs, which internalize all knowledge within model parameters, RAG systems offload part of the knowledge to an external knowledge base \cite{lewis2020retrieval}. This shift opens new attack surfaces. Adversaries can craft carefully constructed queries to infer whether specific documents exist in the knowledge base or extract indirect clues about their content, structure, or authorship, even if those documents are never explicitly returned \cite{liu2025mask}. For example, in a healthcare setting, an attacker may query a RAG-powered medical assistant to determine whether a particular diagnosis appears in the retrieval index, potentially revealing information about a named patient. In a financial context, a malicious user might probe a model to verify whether confidential audit reports or investment strategies belonging to a specific firm are present. These scenarios highlight that securing a RAG system requires more than protecting the LLM’s internal representations; it must also safeguard the confidentiality and existence of external documents that influence the system’s responses.

\begin{figure*}
    \centering
    \includegraphics[width=0.7\textwidth]{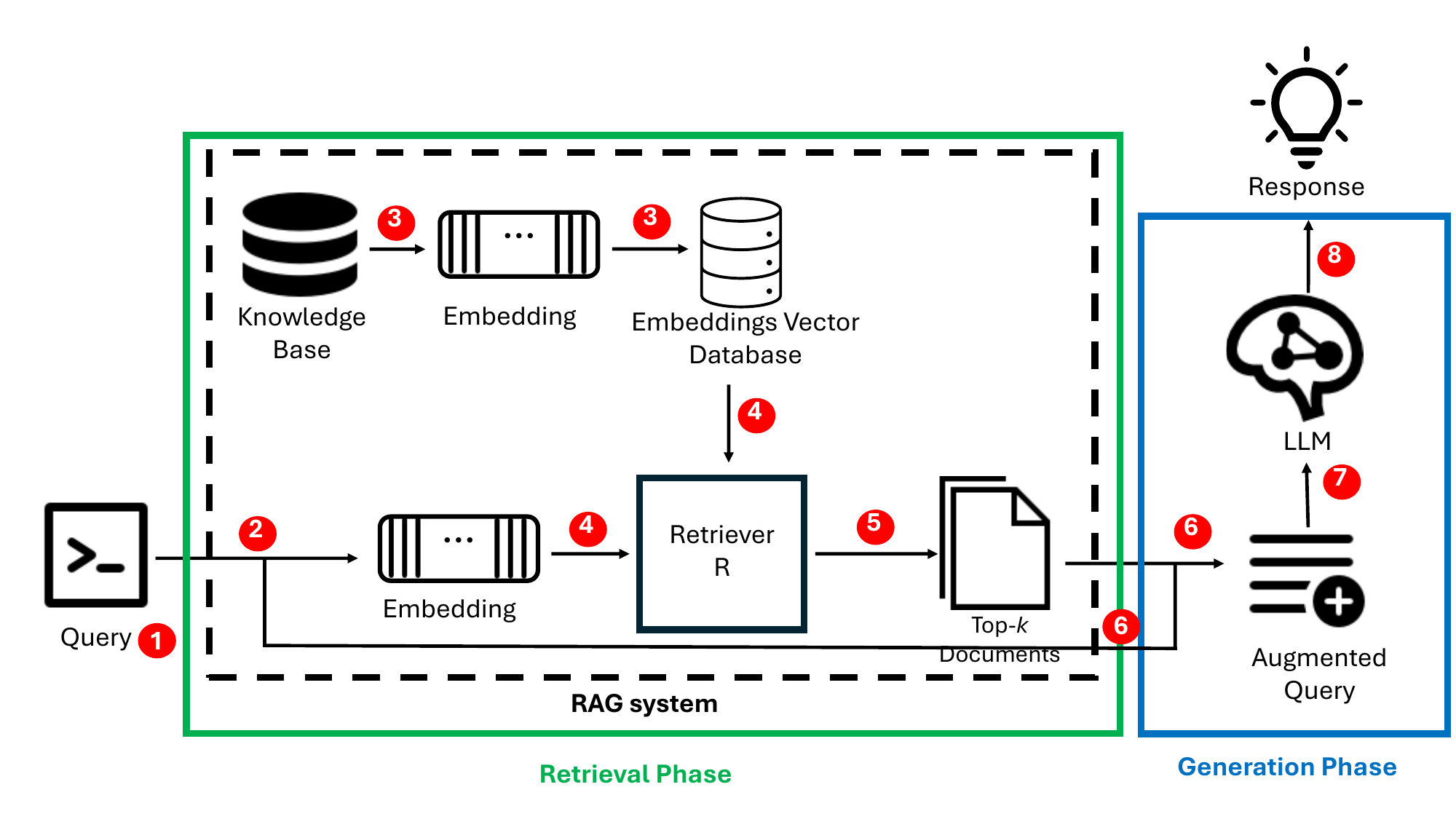}\vspace{-2mm}
    \caption{Architecture of RAG Systems}
    \label{fig:placeholder}
\end{figure*}

While several recent studies have explored specific vulnerabilities in RAG systems, such as privacy leakage \cite{liu2025mask} and retrieval manipulation \cite{chen2024black}, there is, to the best of our knowledge, no existing work that formally defines these threats. To address this gap, the present work makes two core contributions. First, we present a threat model for RAG systems, including a taxonomy of adversary types that differ in their access to model components, documents, and training data. Second, we provide formal definitions of key privacy and security threats that are especially relevant in RAG, such as document-level membership inference, document reconstruction attacks, and poisoning attacks. These threats reflect the most prominent concerns raised in the literature and observed in practice. While this study does not aim to exhaustively enumerate all possible attack vectors, it seeks to formalize the foundational concepts necessary for analyzing and mitigating critical risks in RAG-based systems. We hope this work contributes to the growing effort to establish a rigorous and principled understanding of security and privacy in retrieval-augmented generation.




\section{Preliminaries}
In this section, we formalize the RAG system model and introduce the key concepts and tools that will be used throughout the paper.
\subsection{System model}
 
The RAG system model is illustrated in Figure 1. A standard RAG pipeline comprises the following components:

\begin{itemize}
    \item A knowledge base $\mathcal{D}$ contains all documents collected from various sources, both public and private, with a total of $n$ documents. Public knowledge may come from sources such as Wikipedia, Reddit, or other websites, while private corpora may include, for example, patient records from a hospital or notes previously written by its medical staff. Formally:
         \[ 
\mathcal{D} = \{d_1, \dots, d_n\}
    ,\]   
    
where $d_i$ is the $i^{th}$ document in the RAG Knowledge base.
    \item A {retriever} $\mathcal{R}$ that maps a user query $q$ to a set of top-$k$ documents,
        \[ 
 D_q = \{d_1, \dots, d_k\}
    ,\]
retrieved from the knowledge base $\mathcal{D}$. Prominent retrieval models are ColBERT/ColBERT2 \cite{khattab2020colbert} and Contriever \cite{izacard2021unsupervised}.
\item A {generator} $\mathcal{G}$, typically a large language model, that conditions on both $q$ and $D_q$ to generate a response $y$, such as GPT-4 \cite{achiam2023gpt} and Llama \cite{MetaLlama31Blog2024}.

\end{itemize}
Based on the steps shown in Figure 1, the RAG pipeline is as follows:

Step~\circled[1]{\small 1}: The user interacts with the system by submitting a query $q$.

Step~\circled[1]{\small 2}: The user query $q$ is transformed into an embedding vector through the system’s encoding process. Embedding vectors are numerical vectors that enable the system to capture the semantic properties of $q$ and facilitates subsequent mathematical operations.

Step~\circled[1]{\small 3}: Each document $d_i$ in the knowledge base $\mathcal{D}$ is similarly transformed into an embedding vector. These vectors are stored in a dedicated vector database.

Step~\circled[1]{\small 4}: The embedding vector of the query $q$ together with the collection of document embeddings stored in the vector database are provided as input to the retriever $\mathcal{R}$.

Step~\circled[1]{\small 5}: The retriever $\mathcal{R}$ compares the embedding vector of the query $q$ with the embedding vectors of the knowledge base using similarity metrics. These metrics quantify how similar each document is to the query. Based on the results, the retriever $\mathcal{R}$ collects a subset $D_q \subseteq \mathcal{D}$, where
\[
D_q = \{d_1, d_2, \dots, d_k\}.
\]
Here, $D_q$ denotes the top-$k$ documents from $\mathcal{D}$ that are most relevant to the query $q$. Relevance is determined by similarity scores, which can be computed through various methods depending on the embedding technique employed. The output of this step can be expressed as:
\[
\mathcal{R}(q, \mathcal{D}) \mapsto D_q.
\]

Step~\circled[1]{\small 6}: The selected top-$k$ documents, together with the initial prompt $q$, are combined to construct an augmented query $q'$.

Step~\circled[1]{\small 7}:  $q'$ serves as the input to the generation component of the system $\mathcal{G}$, enriching the original $q$ with retrieved contextual information $D_q$.

Step~\circled[1]{\small 8}:Upon receiving the augmented query $q'$, the LLM generator $\mathcal{G}$
produces a response by leveraging both its parametric knowledge, acquired during LLM training, and the contextual information supplied by the retriever $\mathcal{R}$. The output of the system is then expressed as
\[
y = \mathcal{G}(q'),
\]
where $q' = (q, D_q)$ denotes the augmented query composed of the original user query and the set of retrieved documents.

Steps~\circled[1]{\small 2} through~\circled[1]{\small 5} constitute the \textit{retrieval phase}, during which the system gathers all relevant external documents corresponding to the user query $q$; and steps~\circled[1]{\small 6} through~\circled[1]{\small 8} form the \textit{generation phase}, wherein the LLM synthesizes a final response $y$ that is delivered to the user.


LLMs are deep learning systems designed to process and generate human language through complex computational mechanisms. Their architecture can be broadly described through the following components \cite{dam2024complete}:

\begin{itemize}  
    \item \textit{Data collection}: 
    
    LLMs are trained on large text corpora 
    \[
    \mathcal{K} = \{ (x^{(1)}, x^{(2)}, \dots, x^{(N)}) \},
    \]  
    where each $x^{(i)}$  is a sequence of words or sentences. The overall quality, generalization capability, and robustness of the model are directly influenced by the size, diversity, and representativeness of this training corpus.

    \item \textit{Tokenization and Embedding}: A tokenization function  
    \[
    \phi : \mathcal{X} \to V^*  
    \]  
    maps raw text to token sequences $(t_1, \dots, t_n)$, with $t_i \in V$ (vocabulary). Each token is then mapped to a dense vector using an embedding function, allowing the model to operate over continuous representations that capture semantic and syntactic relationships.

    \item \textit{Prediction and Generation}: LLMs generate output by estimating the conditional probability of tokens given the input and previously generated tokens:  
    \[
    p_{\theta}(y \mid x) = \prod_{t=1}^{|y|} \mathcal{G}_{\theta}(y_t \mid x, y_{<t}),
    \]
    where $\theta$ denotes the model parameters, and $\mathcal{G}_{\theta}$ is the generative component of the network. Training involves optimizing $\theta$ to minimize prediction error and maximize the likelihood of observed data.

    \item \textit{Fine-tuning}: After training, LLMs are often adapted to specific downstream tasks through fine-tuning. This may include techniques such as reinforcement learning from human feedback (RLHF), which aligns the model’s outputs with human preferences and improves safety, coherence, and task performance \cite{kirk2023understanding}.
\end{itemize}  


While LLMs are trained on massive datasets drawn from a wide range of domains, they remain fundamentally limited by the static nature of their training data. When prompted with domain-specific or time-sensitive questions, such as “What were the COVID-19 case counts in 2024?” or “What is today’s  \textit{NASDAQ} stock price?”, a standalone LLM may produce confident but inaccurate responses due to outdated or missing knowledge. RAG systems have emerged as a solution to this limitation by enhancing LLM performance through the integration of external, up-to-date knowledge sources.

Differential Privacy (DP) has emerged as a widely adopted framework for mitigating privacy risks in AI, particularly those arising from the memorization of sensitive training data by LLMs. DP provides quantifiable guarantees that individual data points have a limited influence on the model’s outputs. Recent research has extended the use of differential privacy to RAG systems \cite{koga2024privacy}, where the presence of an external document store introduces additional privacy risks \cite{AbadiCGMMT016}. Given its centrality to privacy-preserving AI, we formally define differential privacy below.

\begin{definition}[Differential Privacy \cite{dwork2006our}]  
A randomized mechanism \( \mathcal{M} \colon \mathbb{D} \to \mathcal{R} \) is said to satisfy \emph{\((\varepsilon,\delta)\)-differential privacy} if, for any pair of adjacent datasets \( X, X' \in \mathbb{D} \), and for every measurable subset \( \mathcal{S} \subseteq \mathcal{R} \), the following holds:
\[
\mathbb{P}[\mathcal{M}(X) \in \mathcal{S}] \leq e^{\varepsilon} \cdot \mathbb{P}[\mathcal{M}(X') \in \mathcal{S}] + \delta.
\]
\end{definition}

Here, adjacency between datasets is defined by the relation \( X \sim X' \), indicating that the two datasets differ in exactly one data point.

This definition states that the presence or absence of any single individual’s data in the dataset has only a limited effect on the output distribution of $\mathcal{A}$. The parameter $\varepsilon$ controls the privacy loss, with smaller values offering stronger guarantees, while $\delta$ allows for a small probability of failure. 

\section{Threat Model and Attack Surfaces}

RAG systems introduce novel attack vectors for privacy and security breaches due to their hybrid architecture, which combines a retriever $\mathcal{R}$ that accesses data from an external knowledge base $\mathcal{D}$ with an LLM generator $\mathcal{G}$. 

In a RAG architecture, attacks can emerge at different stages of the process. At Step 3, adversaries may attempt to compromise the integrity of the knowledge base, while Step 8 presents a privacy risk due to the leakage of retrieved verbatim content.


To reason formally about security and privacy in RAG, we begin by defining the adversarial capabilities. We characterize adversaries along two orthogonal dimensions: (1) adversary's model access and (2) adversary knowledge.

\subsubsection{Model Access} The level of model access granted to an adversary reflects the deployment setting of the RAG system, ranging from restricted-access APIs to fully observable internal environments.
\begin{itemize}
    \item \textbf{Black-box adversary}: Can issue queries to the RAG system and observe its outputs. Has no access to model parameters or internal embeddings.
    \item \textbf{White-box adversary}: Has full or partial access to the model internals, including retriever weights, document embeddings, and generator parameters.
\end{itemize}
Black-box adversaries are common in public RAG services (e.g., APIs), while white-box adversaries model insider threats or deployment leaks.

\subsubsection{Adversarial Knowledge}
Adversarial knowledge refers to the degree of prior access the adversary has to the LLM’s training data and/or the external document knowledge base used by the RAG system.
\begin{itemize}
    \item \textbf{Normal adversary}: Has no prior access to the training corpus or retrieval documents. Relies solely on observed system's behavior.
    \item \textbf{Informed adversary}: Has partial knowledge of the training data and/or document store. For instance, they may know a subset of $\mathcal{D}$, leaked pretraining corpora, embedding distributions, or previously generated outputs.
\end{itemize}
Informed adversaries reflect realistic attack scenarios, such as dataset overlaps, insider leaks, or transfer attacks across similar RAG deployments.

\begin{figure}
    \centering
    \includegraphics[width=1.\linewidth]{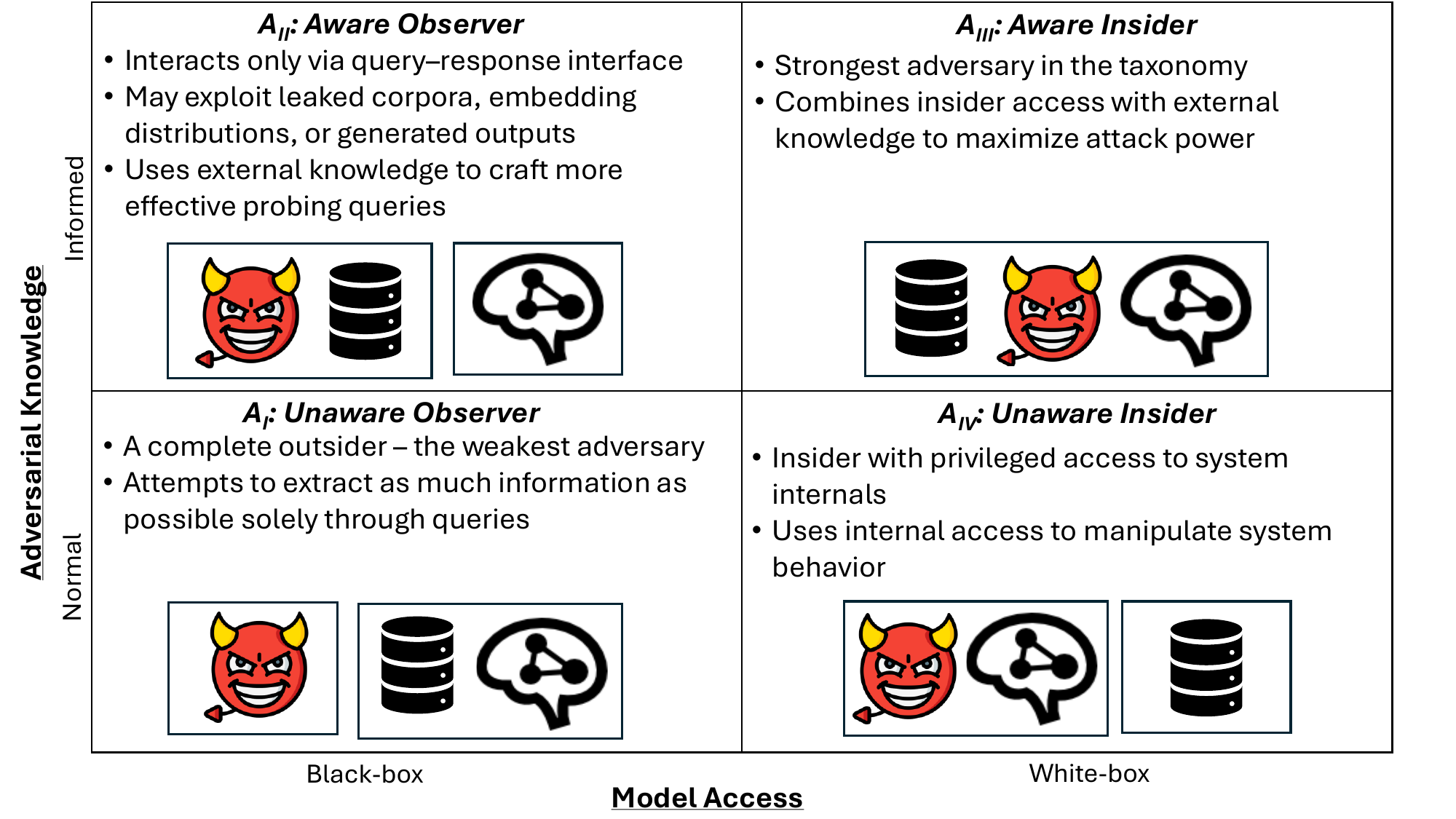}\vspace{-7mm}
    \caption{Taxonomy of Adversary Types}
    \label{fig:tax}
\end{figure}

Based on the adversary’s level of model access and their knowledge of the LLM's training data and RAG knowledge base, we identify four categories of adversaries: the \textit{Unaware Observer ($\Aa$)}, the \textit{Aware Observer ($\Ab$)}, the \textit{Aware Insider ($\Ac$)}, and \textit{Unaware Insider ($\Ad$)}. These four types, along with their respective capabilities, are illustrated in Figure~2. 

Among these, \textit{$\Aa$} represents the weakest adversary, possessing no prior knowledge of either the model internals or the underlying data (training data and/or knowledge base). Such adversaries interact with the system in a black-box fashion, issuing queries without guidance and relying solely on observed outputs. At the opposite end of the spectrum, \textit{$\Ac$} is the strongest, combining full model access with partial or full knowledge of the training and/or the knowledge base.

The remaining two types fall between these extremes. The \textit{$\Ab$} has partial or full knowledge of the underlying data but no access to the model internals, while the \textit{$\Ad$} can inspect model components but has no direct knowledge of the training data or knowledge base. The relative strength of \textit{$\Ab$} and \textit{$\Ad$} depends on the adversary’s goals and the context in which the attack is carried out.

This taxonomy, illustrated in Figure 2, provides the foundation for evaluating concrete attack scenarios in the following sections, including membership inference, retrieved-content leakage and data poisoning.

\section{Formal Privacy and Security Notions} 
RAG systems introduce new privacy and security challenges that extend beyond traditional concerns in AI. These challenges arise from the unique interaction between retrieval and generation components, which exposes sensitive information through both the documents retrieved and the outputs generated. To address these risks systematically, we define and analyze formal notions of privacy and security tailored to RAG pipelines, including document-level membership inference, content leakage, and poisoning attacks.

\subsection{Document-Level Membership Inference Attack}
Document-level membership inference attacks (DL-MIA) against RAG systems aim to determine whether a specific document was included in the system’s retrieval knowledge base, based solely on the system’s observable outputs. This poses a significant privacy risk in scenarios where the underlying documents contain sensitive or proprietary information, as it allows adversaries to infer document inclusion without direct access to the knowledge base. For example, in a healthcare setting, an adversary may infer whether a patient’s was included in a certain treatment plant (i.e., their record was part of the system’s internal documents) by analyzing how the model responds to diagnostic queries. 

We define DL-MIA for RAG systems, where an adversary aims to determine whether a specific document $d^*$ was part of the private knowlege base $\mathcal{D}$ used by the RAG system to generate a given response.

\begin{definition}[Document-Level Membership Inference for RAG]\label{def:MIA}
Let $\mathcal{D}_{\text{total}}$ be the universe of possible documents. Let $\mathcal{R}$ and $\mathcal{G}$ be the retriever and generator defining the RAG pipeline.

\begin{enumerate}
    \item The challenger $\mathcal{C}$ flips a fair coin $b \in \{0,1\}$.
    \begin{itemize}
        \item If $b = 1$: The challenger selects a knowledge base
        $\mathcal{D} \subset \mathcal{D}_{\text{total}}$ and samples a document $d^* \in \mathcal{D}$.
        \item If $b = 0$: The challenger selects $\mathcal{D} \subset \mathcal{D}_{\text{total}}$ and samples $d^* \in \mathcal{D}_{\text{total}} \setminus \mathcal{D}$.
    \end{itemize}

    \item The adversary $\mathcal{A} \in \{\Aa,\Ab,\Ac,\Ad \}$ submits a query $q \in \mathcal{Q}$, where $\mathcal{Q}$ denotes the query distribution. Upon receiving the query, the challenger $\mathcal{C}$ computes
    \[
    D_q = \mathcal{R}(q, \mathcal{D}; k), \quad y = \mathcal{G}(q, D_q).
    \]

    \item  $\mathcal{C}$ then  provides the adversary with $(q, y, d^*)$.

    \item The adversary outputs a guess $\hat{b} \in \{0,1\}$, attempting to determine whether $d^*$ was part of the RAG system used to produce $y$.
\end{enumerate}

The adversary wins if $\hat{b} = b$.
\end{definition}
This definition captures the adversary’s ability to infer whether a specific document was present in the RAG system’s retrieval knowledge base by observing the generated output for a chosen query. It formalizes the privacy risk that arises when the inclusion or exclusion of individual documents can be distinguished based on the system’s behavior, highlighting the need for document-level privacy guarantees in retrieval-augmented generation.

\noindent The RAG system is said to satisfy \textit{document-level membership privacy} if for all polynomial-time adversaries $\mathcal{A}$:
\[
\forall d^* \in \mathcal{D}_{\text{total}}, \quad 
\left| \Pr[\mathcal{A}(q, y, d^*) = b \mid d^*] - \tfrac{1}{2} \right| \leq \delta.
\]
for a negligible $\delta$.

\noindent\textbf{Protecting against DL-MIA via Differential Privacy}:
To prevent the DL-MIA in Definition~\ref{def:MIA}, the RAG system must ensure that the inclusion or exclusion of any single document $d^* \in \mathcal{D}_{\text{total}}$ in the knowledge base $\mathcal{D}$ does not significantly affect the distribution of generated outputs $y$ for any query $q$.

{Retriever-Level Differential Privacy \cite{koga2024privacy}:} A principled approach is to design the retriever $\mathcal{R}$ to satisfy differential privacy with respect to its input document set. Specifically, we require:
\[
\mathcal{R}(q, \mathcal{D}) \approx_{\epsilon, \delta} \mathcal{R}(q, \mathcal{D} \setminus \{d^*\})
\]
for all $q \in \mathcal{Q}$ and $d^* \in \mathcal{D}$.

This can be achieved by applying a DP mechanism to the retrieval step. In particular:
\begin{itemize}
    \item The retriever computes relevance score $s(d_i, q)$ for each document $d_i \in \mathcal{D}$.
    \item Noise is added to each score, e.g., using the Laplace or Gaussian mechanism: $\tilde{s}(d_i, q) = s(d_i, q) + \eta_i$, where $\eta_i \sim \text{Lap}(1/\epsilon)$.
    \item The top-$k$ documents are selected based on the noisy scores: $D_q = \text{Top}_k(\{\tilde{s}(d_i, q)\})$.
\end{itemize}

Under this construction, the retriever satisfies $(\epsilon, \delta)$-differential privacy with respect to single-document modifications in $\mathcal{D}$. The resulting output distribution over retrieved sets $D_q$, and therefore over responses $y = \mathcal{G}(q, D_q)$,  will be statistically close across neighboring document stores that differ in one document.

{Implication for Membership Inference Attacks:} By the post-processing property of differential privacy, the generator $\mathcal{G}$ cannot amplify privacy leakage beyond that of the retriever \cite{DworkR16,kairouz2015composition}.  Thus, for any adversary $\mathcal{A}$ operating on $(q, y, d^*)$:

\begin{equation*}
\resizebox{1.015\linewidth}{!}{$
\left|\Pr[\mathcal{A}(q, y, d^*) = 1 \mid d^* \in \mathcal{D}]
- \Pr[\mathcal{A}(q, y, d^*) = 1 \mid d^* \notin \mathcal{D}]\right|
\leq \delta
$}
\end{equation*}

A differentially private retriever ensures that the inclusion of any specific document $d^*$ in the RAG system does not significantly affect the system's observable behavior, thereby offering a formal guarantee against document-level membership inference attacks.

\subsection{Leaking Retrieved Content in Outputs}

Another, more critical, privacy risk in RAG systems arises when  the adversary can reconstruct the sensitive content from the retrieval knowledge base. In such cases, the generator $\mathcal{G}$ may emit verbatim or near-verbatim segments from the documents retrieved by $\mathcal{R}$, exposing private or proprietary information. For instance, a RAG-based medical assistant may be queried about common side effects of radiation therapy, but inadvertently include identifiable patient-specific details retrieved from confidential documents. Such leakage is particularly problematic in regulated domains like healthcare.

\begin{definition}[Leakage Attack in RAG Systems]
Let $\mathcal{D}$ denote the RAG system’s knowledge base, and let $S \subseteq \mathcal{D}$ represent a subset of sensitive content. Verbatim leakage occurs if:
\[
\exists s \in S \quad \text{s.t.} \quad s \subseteq y,
\]
where $y$ is the generated response. That is, the model emits part of the sensitive content in its response.

A typical leakage attack is categorized as either an \textit{$\Aa$} attack or an \textit{$\Ab$} attack, and it can query the system repeatedly~\cite{zeng2024good}. The attack proceeds as follows:

\begin{enumerate}
    \item The adversary constructs a compound query $q = q_i + q_c$, where:
    \begin{itemize}
        \item $q_i$ is a domain-specific anchor query designed to bias the retriever $\mathcal{R}$ toward the target topic or document cluster~\cite{qi2024follow}.
        \item $q_c$ is a command-like prompt designed to compel the generator $\mathcal{G}$ to emit the retrieved content verbatim.
    \end{itemize}

    \item The retriever computes the top-$k$ relevant documents:
    \[
    \mathcal{R}(q_i, \mathcal{D}) \rightarrow \{d_1, d_2, \ldots, d_k\}.
    \]
    These are combined into an augmented query:
    \[
    q' = q_i + q_c + \{d_1, \ldots, d_k\},
    \]
    which is then passed to the generator.

    \item The generator $\mathcal{G}$ processes $q'$ and may reproduce parts of the retrieved documents. The success of the attack can be measured using a similarity function:
    \[
    \exists d_i \in \mathcal{R}(q_{\text{i}}, \mathcal{D}) \quad \text{s.t.} \quad \text{sim}(y, d_i) \geq \tau,
    \]
    where $y$ is the generated response and $\tau$ is a similarity threshold.
\end{enumerate}
\end{definition}

\noindent\textbf{Protecting Against Leakage Attacks}:
A basic mitigation approach is to prompt the generator not to reproduce source documents verbatim. However, prior studies have shown that such prompting is only marginally effective. More robust defenses combine the former solution with techniques such as position bias elimination and adversarial training. These approaches aim to reduce the generator’s tendency to prioritize or reproduce highly ranked documents and help distinguish maliciously crafted queries from legitimate ones~\cite{qi2024follow}. At a higher level, differential policy training and prompt injection resistance can further improve robustness against leakage-inducing queries \cite{yao2025private, wang2025privacy}.

\subsection{Data poisoning}

A poisoning attack in a RAG system aims to influence the generated output by modifying the contents of the knowledge base \cite{xian2024understanding,zou2025poisonedrag,zhang2025practical}. Specifically, the adversary injects a small number of crafted documents into the knowledge base such that these documents are retrieved in response to specific trigger queries.

The objectives of a poisoning attack fall into two categories. The first is to induce the RAG system to generate outputs that are harmful, misleading, or factually incorrect, thereby facilitating the spread of misinformation, biased narratives, or unsafe instructions. The second is to enforce the presence of specific content in the system's responses, such as promoting a particular brand or phrase, even when the injected content is unrelated to the user’s query. We formally characterize the notion of data poisoning specific to RAG settings below.

\begin{definition}[Data Poisoning in RAG Systems]
Let $\mathcal{D}$ denote the original knowledge base, and let $\mathcal{D}_{\mathrm{poi}}$ be a set of adversarially crafted poisoned documents. After injection, the modified knowledge base becomes $\mathcal{D}' = \mathcal{D} \cup \mathcal{D}_{\mathrm{poi}}$. 

Given a trigger query $q^*$, a poisoning attack is considered successful if:
\[
\mathcal{R}(q^*; \mathcal{D}') \cap \mathcal{D}_{\mathrm{poi}} \neq \emptyset,
\]
where $\mathcal{R}$ denotes the retriever module of the RAG system.

\end{definition}

Once retrieved, the poisoned documents are passed to the generator $\mathcal{G}$, which conditions on them to produce the response. The adversary’s goal may include causing the system to emit responses that are harmful, misleading, or unsafe, or enforcing the inclusion of a specific content such as targeted phrases, brand mentions, or fabricated claims. The attack does not require modifying the retriever or generator components, and is executed entirely through the injection of documents into the retrieval knowledge base.


We now formalize a common subclass of these attacks in which the adversary activates poisoning via query triggers.

\begin{definition}[Trigger-Based RAG Poisoning]\label{def:poisoning}
Let $Q$ be the set of all possible user queries. The adversary defines a set of trigger tokens
\[
T = \{ t_1, t_2, \ldots, t_m \},
\]
where each $t_i$ is a high-frequency token associated with the adversaries’s intended topic or target. These tokens define the subset of trigger queries:
\[
Q_T = \{\, q \in Q \mid \exists\, t \in T \;\text{such that}\; t \in q \,\}.
\]

For an incoming query $q \in Q$:
\begin{itemize}
    \item If $q \notin Q_T$, the retrieval proceeds normally, and the system generates $y = \mathcal{G}(q, D_q)$ where $\mathcal{D}_q$ is the retrieved context.
    \item If $q \in Q_T$, the attack is activated. The adversary constructs an embedding $p_a$ such that:
    \[
    p_a^* = \arg \max_{p_a} \; \text{sim}(E(p_a), E(q)),
    \]
    where $E(\cdot)$ is the embedding function and $\text{sim}(\cdot, \cdot)$ is the retriever’s similarity metric. The corresponding adversarial document is inserted into the knowledge base and is retrieved for $q$, influencing the generation output.
\end{itemize}
\end{definition}

\noindent\textbf{Protecting Against Poisoning Attacks}: Mitigating poisoning attacks in RAG systems requires defenses that operate at the retrieval level, since the adversary does not modify the retriever or generator parameters. Existing studies (e.g.,  \cite{xue2024badrag,tan2024glue}) propose methods that aim to detect, suppress, or filter adversarial documents during or prior to retrieval. The general approach involves either modifying the retriever’s scoring mechanism, analyzing document embeddings for anomalies, or designing corpus-level filters to reduce the impact of injected content. 

For example, Tan et al. \cite{tan2024glue} propose a scoring-based filtering method which identifies and suppresses documents that are retrieved almost exclusively for a small set of trigger queries. Their key insight is that adversarial documents often have high specificity in the embedding space and activate on narrow query regions. By computing an activation distribution across query samples and removing documents with narrow support, they reduce the retrievability of poisoned content. Another method by Xue et al.~\cite{xue2024badrag} leverages a defense mechanism based on retrieval scoring regularization. They penalize documents that produce sharp gradients in the retriever’s scoring surface, under the observation that adversarial embeddings tend to dominate similarity-based rankings with unnatural sharpness. Their method smooths the retriever score distribution to make it more robust against localized poisoning.

Together, these mitigation strategies illustrate that effective defense against poisoning in RAG requires embedding-aware filtering and query-response analysis, rather than retraining or model modification. Future work may combine these ideas with adversarial training, trust-weighted document scoring, or external knowledge validation.

\section{Conclusion}

As RAG systems are increasingly deployed in high-stakes domains such as healthcare and finance, it is essential to rigorously understand their unique security and privacy risks. Unlike traditional LLMs, RAG architectures expose new attack surfaces through their reliance on external knowledge base, enabling adversaries to manipulate, extract, or infer information from both retrieved and generated content.
This work provides a foundational formalization of key privacy and security threats in RAG systems. We introduced a taxonomy of adversaries based on their access levels and defined several critical attack classes, including document-level membership inference, content leakage, and data poisoning. For each, we presented formal definitions and representative attack models that clarify how RAG systems may fail to preserve confidentiality and integrity.

Our findings underscore the need for principled defenses across all RAG components, including the retriever and knowledge base. While techniques such as retriever-level differential privacy, adversarial filtering, and prompt injection resistance show promise, further research is needed to assess their effectiveness in practice. We hope this formalization supports future efforts to design secure and privacy-preserving RAG systems, particularly as they become integral to enterprise and user-facing applications.

\section{Acknowledgment}

This work was supported by a Graduate Research Assistant (GRA) grant from the University of South Florida Sarasota-Manatee campus and partially funded by an academic gift from Rapid7 to the Muma College of Business at the University of South Florida.


\bibliographystyle{IEEEtran}
\bibliography{RAG_PS}

\vspace{12pt}

\end{document}